\documentclass[showpacs,preprintnumbers,amsmath,amssymb,apl,twocolumn]{revtex4}
\usepackage{amsfonts}
\usepackage{mathrsfs}
\usepackage{graphicx}
\usepackage{dcolumn}
\usepackage{bm}

\begin{document}

\title{Solid-State Optimal Phase-Covariant Quantum Cloning Machine}

\author{Xin-Yu Pan\footnote{xypan@aphy.iphy.ac.cn}, Gang-Qin Liu, Li-Li Yang,
Heng Fan\footnote{hfan@iphy.ac.cn}}
\affiliation{%
Institute of Physics, Chinese Academy of Sciences, Beijing
100190, China
}%
\date{\today}

\begin{abstract}
Here we report an experimental realization of optimal
phase-covariant quantum cloning machine with a single electron spin
in solid state system at room temperature. The involved three states
of two logic qubits are encoded physically in three levels of a
single electron spin with two Zeeman sub-levels at a
nitrogen-vacancy defect center in diamond. The preparation of input
state and the phase-covariant quantum cloning transformation are
controlled by two independent microwave fields. The average
experimental fidelity reaches $85.2\% $ which is very close to
theoretical optimal fidelity $85.4\%$ and is beyond the bound
$83.3\%$ of universal cloning.

\end{abstract}

\maketitle

Nitrogen-vacancy (NV) defect center in diamond is one of the most
promising systems to be the solid state quantum information
processors \cite{Gruber97,Neumann08}. It can be individually
addressed, optically polarized and detected, and is with excellent
coherence properties. Both electronic and nuclear spins at the NV
centers can be well controlled. The advantages of the NV centers for
quantum information processing are their scalability, and their long coherence time $T_2$ at room
temperature, which can be further prolonged
\cite{dynamicaldecoup,dynamicaldecoup2,dynamicaldecoup3,Naydenovpreprint10}.
Despite its scalability,
an individual electronic spin at NV center in diamond is still very
useful, such as for real applications and being a test bed for
quantum algorithms
\cite{Mazelukinnature,Taylornaturephys,Shiduprl10,Toganlukinnature10}.

In this Letter, with a coherent superposition of all three levels of
a single electronic spin, we demonstrate the optimal phase-covariant
quantum cloning.

It is well known that a quantum state can not be cloned \cite{WZ}.
However, we can try to clone a quantum state approximately or
probabilistically, see for example \cite{BH,cv,du11}.
The no-cloning theorem is fundamental
for the security of the quantum key distribution protocols in
quantum cryptography, for example for the well-known BB84 protocol
\cite{BB84}.
The optimal cloning machine for BB84 states is the phase-covariant
quantum cloning machine \cite{BCDM,FMWW,Du,Chendupra07} for which
the input state is in a specified form $|\psi \rangle =(|0\rangle +
e^{i \phi }|1\rangle )/\sqrt {2}$, i.e., all input states are
located in the equator of the Bloch sphere, see FIG. 1(a). The
fidelity of the phase-covariant quantum cloning machine is around
$85.4\% $ which is better than around $83.3\%  $ of the optimal
universal cloning.

\begin{figure}
\includegraphics[width=0.38\textwidth]{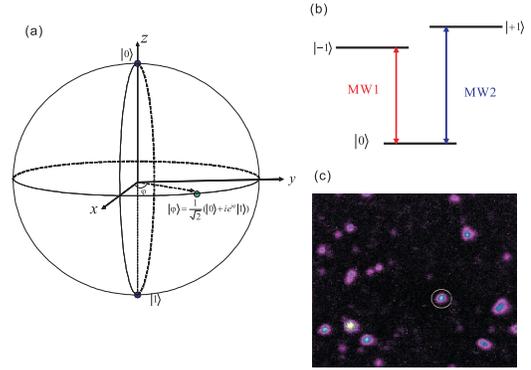}
\caption{(color online) Bloch sphere and energy level for nitrogen
vacancy center in diamond.(a) The states need to be cloned are in a
specified form which are located in the equator of the Bloch sphere
$|\psi \rangle =(|0\rangle + e^{i \phi }|1\rangle )/\sqrt {2}$. (b)
Energy level of the NV center in diamond. (c) Two-dimension scanning
confocal image of the sample. Bright spot circled is the NV center
we investigate.}
\end{figure}

A NV center comprises a substitutional nitrogen atom instead of a
carbon atom and an adjacent lattice vacancy. Experiments are carried
out in a type Ib diamond nanocrystal from company Element Six. The
average size of diamond nanocrystal is about 30 nm. Single NV
defects in diamond are addressed by a home-built laser scanning
confocal microscope system at room temperature [FIG. 1(c)]. A 532 nm
continuous-wave laser modified by an acoustic optical modulator
(AOM) with a rising edge of 10 ns is focused onto the sample with a
microscope objective(numerical aperture=0.9). The fluorescence is
also collected by the same objective, and passes through a 650 nm
long-pass filter. Fluorescence signal is detected by a single photon
counting module (SPCM, Perkin-Elmer) with a National Instruments
counter 6602. Second order photon correlation function $g^2(\tau)$
of center A indicates that it is a single quantum emitter [FIG.
2(a)].

\begin{figure}
\includegraphics[width=0.38\textwidth]{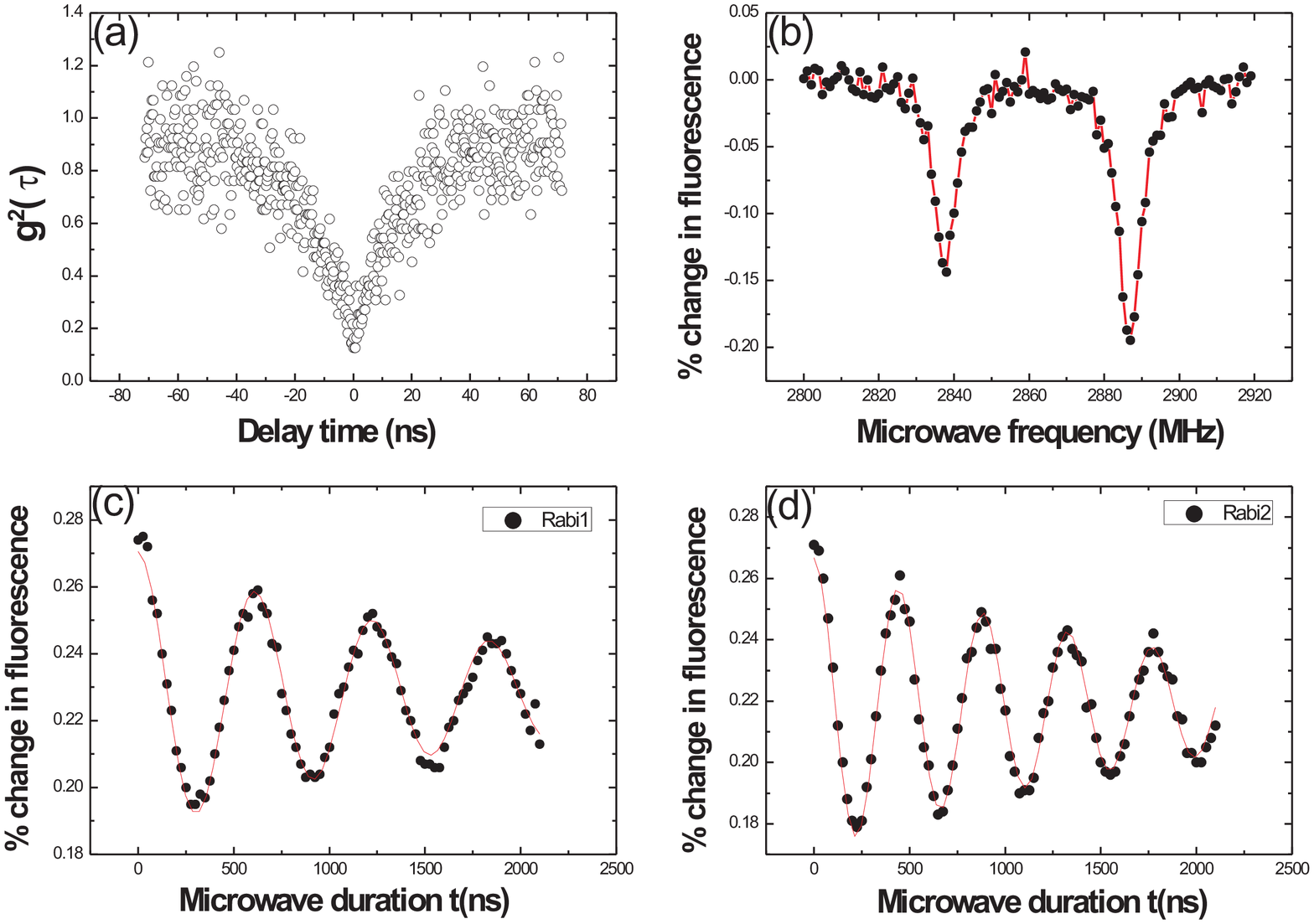}
\caption{(color online) Second order photon correlation function,
ESR spectrum, Rabi oscillations of two transitions.  (a) Second
order photon correlation function $g^2(\tau)$ of the NV center. (b)
ESR spectrum of the NV center. Two main peaks correspond to
\emph{m$_{s}$}=1 and \emph{m$_{s}$}=-1. (c) Rabi oscillations for
the transition between \emph{m$_{s}$}=0 and \emph{m$_{s}$}=1. (d)
Rabi oscillations for the transition between \emph{m$_{s}$}=0 and
\emph{m$_{s}$}=-1.}
\end{figure}

The Hamiltonian with electron spin zero
field splitting and the electron
Zeeman interaction takes the form,
\begin{eqnarray}
H= \textbf{S}\bar {D}\textbf{S}+\beta _e\vec{B}_0\bar
{g}_e\textbf{S},
\end{eqnarray}
where $g_e$ and $\beta _e$ are
the $g$ factor and Bohr
magneton for electron, $\vec {B}_0$ is the applied magnetic field.

Experimentally, a microwave radiation is sent out by a copper wire
of 20 $\mu$m diameter placed with a distance of 20 $\mu$m from the
NV center. The Electron Spin Resonance (ESR) spectrum is shown in
FIG. 2(b) as a function of the fluorescence change against the
microwave frequency without external magnetic field, this is due to
symmetry breaking of this NV center corresponding to a non-zero
magnetic field. The two resonant frequencies
correspond to the transitions of \emph{m$_{s}$}=0 to
\emph{m$_{s}$}=1 and \emph{m$_{s}$}=0 to \emph{m$_{s}$}=-1. We
denote the corresponding states as $|m_{s}=0\rangle _p$ and
$|m_{s}=\pm 1 \rangle _p$, where the subindex $p$ means those states
are physical states to differ them from the logic qubits. In our
experiment, the cloning processing is to transfer state $|\psi
\rangle |0\rangle $ to two copies $|output \rangle =\frac {1}{\sqrt
{2}}|00\rangle +\frac {1}{2}e^{i \phi }|01\rangle +\frac {1}{2}e^{i
\phi }|10\rangle $. We use the encoding scheme: $-i|00\rangle \sim
|m_s=-1\rangle _p;|10\rangle \sim |m_s=0\rangle _p;-i|01\rangle \sim
|m_s=1\rangle _p$.

To control the electron spin state,
first, a laser pulse initializes the spin state to $|m_s=0\rangle
_p$; then the microwave pulses of weak power are used to manipulate
the spin state; finally, the spin state is read out by the
fluorescence intensity under a second laser excitation. The Rabi
oscillation of the electron spin of single NV center is shown in
FIG. 2(c) and (d), the scatter points are experiment data and each
point is a statistical average result typically repeated $10^5$
times, the red line is the fitting using a function of a cosine with
an exponential decay.

\begin{figure}
\includegraphics[width=0.38\textwidth]{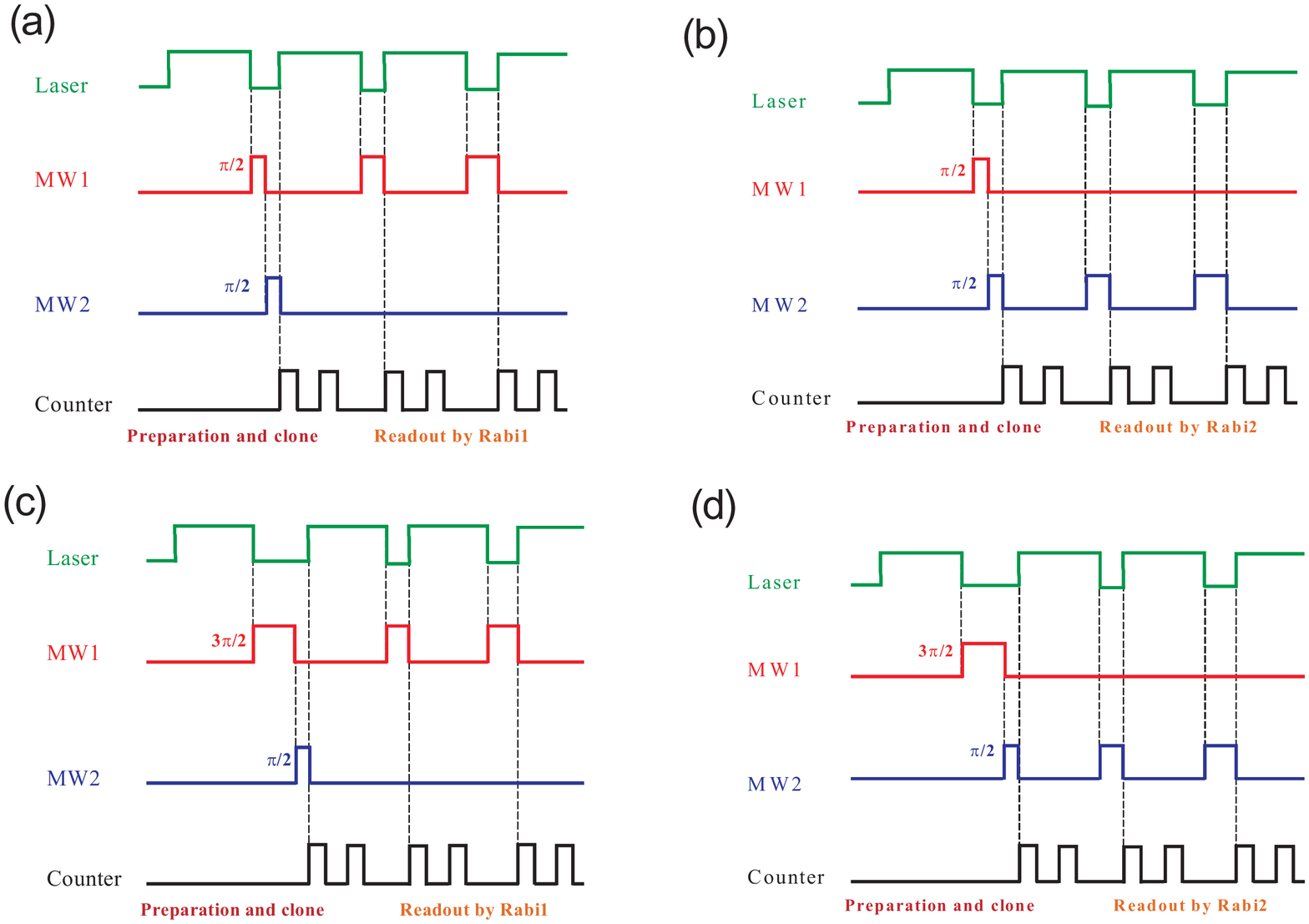}
\caption{(color online)Scheme for quantum phase cloning.  (a)A MW1
$\pi /2$ pulse creates $(|m_s=0\rangle _p+ i|m_s=-1\rangle _p)/\sqrt
{2}$ state, then apply MW2 for another $\pi /2$ pulse for quantum
phase cloning.  After all we measure the standard Rabi oscillations
for transition between $|m_s=0\rangle _p$ and $|m_s=-1\rangle _p$.
(b) The same pulse sequence for the phase cloning, but we measure
the Rabi oscillations for transition between $|m_s=0\rangle _p$ and
$|m_s=1\rangle _p$. (c) A MW1 $3\pi /2$ pulse can create
$(|m_s=0\rangle _p- i|m_s=-1\rangle _p)/\sqrt {2}$ state, after
phase cloning we measure Rabi oscillations with MW1. (d) The MW1
$3\pi /2$ pulse create $(|m_s=0\rangle _p- i|m_s=-1\rangle _p)/\sqrt
{2}$ state, after cloning we measure Rabi oscillations with MW2.}
\end{figure}

FIG. 3 shows the scheme for quantum phase cloning.
The output state should be a superposition state $|output\rangle
_p=\frac {1}{2}|m_s=0\rangle _p+i\frac {1}{2}|m_s=1\rangle _p+
i\frac {1}{\sqrt {2}}|m_s=-1\rangle _p.$
The scheme for measure is by MW1 to confirm $|output\rangle _p$ is
superposed by a pure state $\frac {1}{2}|m_s=0\rangle _p+ i\frac
{1}{\sqrt {2}}|m_s=-1\rangle _p$, and by MW2 to confirm the pure
state form $\frac {1}{2}|m_s=0\rangle _p+i\frac {1}{2}|m_s=1\rangle
_p$.
Combination of those measured results indicate that the output is in
form $|output\rangle _p$.
By analyzing the experiment data, the exact form of the
output state and the fidelity can be obtained.
We then can repeat those experimental steps except that with
different state preparation.
The experimental
results are shown in FIG. 4.

\begin{figure}
\includegraphics[width=0.38\textwidth]{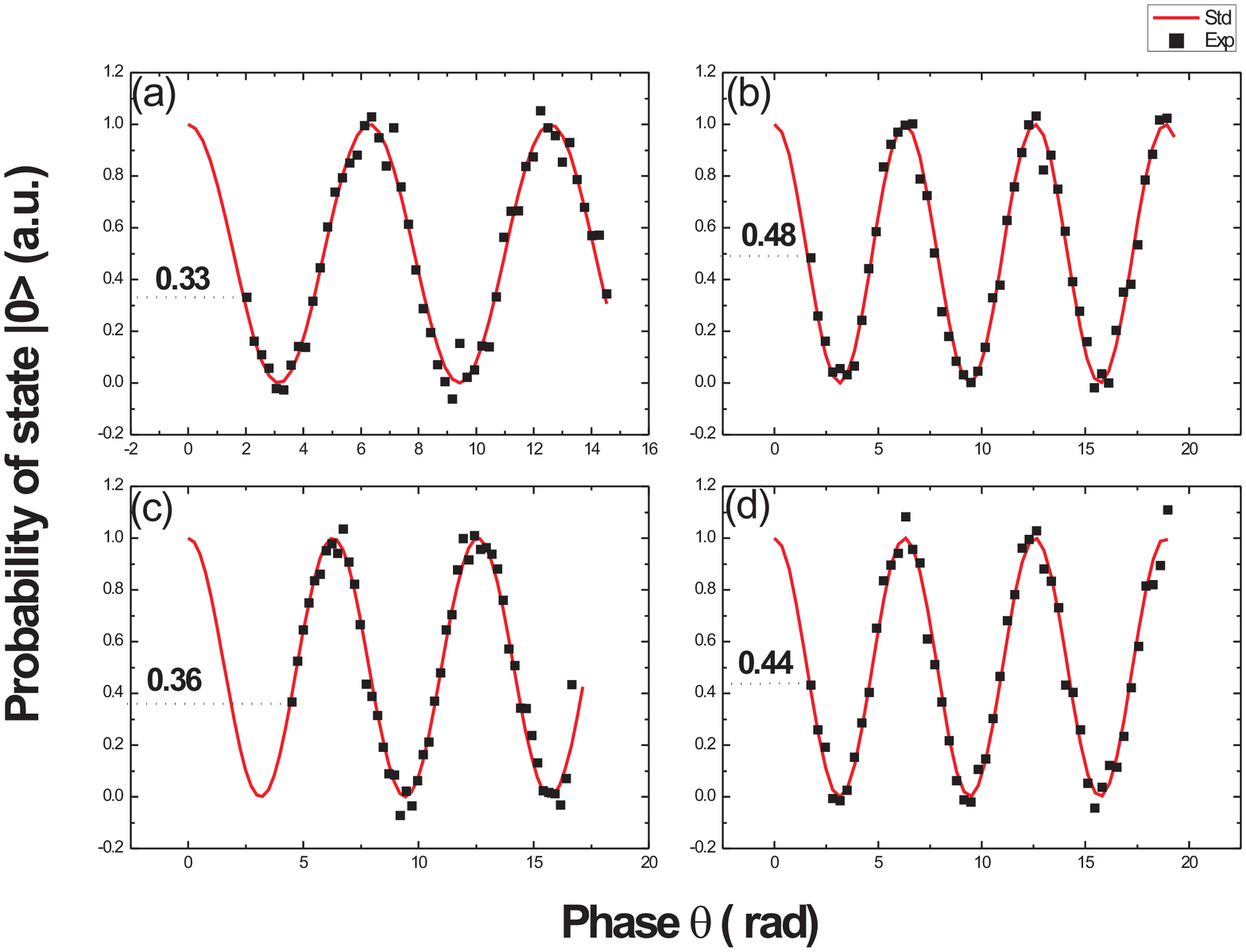}
\caption{(color online)Measured results of the quantum phase
cloning. (a) Red line is the standard dependence of probability of
the state $|m_s=0\rangle _p$ on the phase of microwave pulse, by
applying pulse sequence Figure 3(a), the black square is the
experiment results for Rabi oscillations of transition with MW1, the
start point of this curve determines the population probability at
the state $|m_s=0\rangle _p$ is 33 \%, theory is also around 33\%.
(b) Start point of the curve determines the probability at the state
$|m_s=0\rangle _p$ is 48\%, theory is 50\% . (c) and (d) The start
points of the curves determines the probability at the state
$|m_s=0\rangle _p$ are 36\% and 44\%, theory are 33\% and 50\% . }
\end{figure}

The measured data by MW1 shows clearly Rabi oscillation which
represents that the state of NV is in a superposed state. Also from
the start point of Rabi oscillation,
$\alpha $, the relative rate of fluorescence, we know that the
measured state is in form $\sqrt {\alpha }e^{i\phi }|m_s=0\rangle _p
+i\sqrt {1-\alpha }|m_s=-1\rangle _p$. Similarly with starting rate
of fluorescence $\beta $ of MW2, we know the state is $\sqrt {\beta
}|m_s=0\rangle _p +i\sqrt {1-\beta }e^{i\phi }|m_s=1\rangle _p$. The
combination of those two results show that the NV center state
should be the form
\begin{eqnarray}
&&\sqrt {\alpha \beta }e^{i\phi }|m_s=0\rangle _p +i\sqrt {(1-\alpha
)\beta }|m_s=-1\rangle _p \nonumber \\
&&~~~ +i\sqrt {(1-\beta )\alpha }e^{i\phi }|m_s=1\rangle _p
\label{output}
\end{eqnarray}
with normalization $1/\sqrt {\alpha +\beta -\alpha \beta }$.
Using the experimental data in FIG. 4(a,b),
we find two fidelities are $F_1=84.6\% $ and $F_2=86.1\% $, both are
beyond the bound of the optimal fidelity of universal quantum
cloning. By data in FIG. 4(c,d), we find the two fidelities of the
two copies in output are $F_1=82.9\%$ and $F_2=87.1\%$,
we find $F_1+F_2-\sqrt {(1-F_1)(1-F_2)}\approx 1.55$, which clearly
larger than $1.5$ of the universal cloning \cite{Cerf}. Thus phase
cloning is better than the universal case.
By average, we find the
experimental cloning fidelity reaches $85.2\%$ which is very close
to theoretical bound $85.4\%$ and apparently beyond the bound of the
universal quantum cloning.

\begin{figure}
\includegraphics[width=0.4\textwidth]{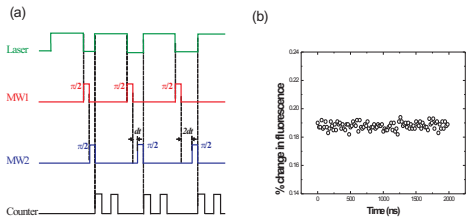}
\caption{(color online) Quantum phase cloning for input state with
different phases. (a) MW1 $\pi /2$ pulse is applied to create
$(|m_s=0\rangle _p+ i|m_s=-1\rangle _p)/\sqrt {2}$ state, before
applying MW2, we wait for time $jdt, j=1,2,...,dt=20ns,50ns$ so that
state $(|m_s=0\rangle _p+ i|m_s=-1\rangle _p)/\sqrt {2}$ evolves
freely to another state $(|m_s=0\rangle _p+ e^{i \omega jdt
}i|m_s=-1\rangle _p)/\sqrt {2}$ with additional relative phase
$\omega jdt$ depending on waiting time $jdt$ and rotating speed $\omega $ determined
by environment, see
\cite{thesis}. (b) Experiment results show that with different waiting time
periods within time scale 2 $\mu $s, after phase cloning operation,
the intensity of fluorescence of the output state is stable which
agrees with theory expectation.}
\end{figure}

A phase quantum cloning need the input state with an arbitrary
phase. Experimental procedures are shown in FIG.5.
This finishes the implementation of the whole quantum phase cloning.

In summary, we report the solid-state phase-covariant quantum
cloning machine implementation in experiment at room temperature.
Our observation shows that two microwave fields MW1 and
MW2 can be combined to create an arbitrary superposition three-level
state in quantum phase cloning processing and for other aims.
This
can be used as a basis for scalable, precisely controllable and
measurable three-level quantum information devices.

This work is supported by NSFC (10974247, 10974251) and ``973''
programs (2009CB929103, 2010CB922904).

\end{document}